\documentstyle[psfig]{article}
\newcommand{\be}{\begin{equation}}
\newcommand{\ee}{\end{equation}}
\newcommand{\beqas}{\begin{eqnarray*}}
\newcommand{\eeqas}{\end{eqnarray*}}
\newcommand{\beqar}{\begin{eqnarray}}
\newcommand{\eeqar}{\end{eqnarray}}

\begin{document}
\title{Power-laws in economy and finance: some ideas from physics}

\author{Jean-Philippe
Bouchaud$^{\dagger,*}$}
\date{{\small $^\dagger$ Science \& Finance, 109-111 rue Victor Hugo,
92532
Levallois {\sc cedex}, FRANCE;\\ http://www.science-finance.fr\\
$^*$ Service de Physique de l'\'Etat Condens\'e,
 Centre d'\'etudes de Saclay,\\
Orme des Merisiers,
91191 Gif-sur-Yvette {\sc cedex}, FRANCE\\}
\today}
\maketitle

\begin{abstract}
We discuss several models in order to shed light on the origin of power-law {\it distributions} 
and power-law {\it correlations} in financial time series. From an empirical point of view, the exponents describing
the tails of the price increments distribution and the decay of the volatility 
correlations are rather robust and suggest universality. However, many of
the models that appear naturally (for example, to account for the distribution of wealth) 
contain some multiplicative noise, which generically leads
to {\it non universal exponents}. Recent progress in the empirical study of the volatility 
suggests that the volatility results from some sort of multiplicative cascade. A convincing 
`microscopic' (i.e. trader based) model that explains this observation is however not yet available. It would be particularly important to understand the
relevance of the pseudo-geometric progression of natural human time scales on the long range nature of the volatility correlations. 

\end{abstract}

\section{Introduction}

Physicists are often fascinated by power-laws. The reason for this is that
complex, collective phenomenon do give rise to power-laws which are
{\it universal}, that is to a large degree independent of the microscopic 
details of the phenomenon. These power-laws emerge from collective 
action
and transcends individual specificities. As such, they are unforgeable 
signatures of a collective mechanism. Examples in the physics literature 
are numerous.
A well known example is that of phase transitions, where a system evolves 
from a disordered state to an ordered state: many observables behave as
 universal power laws in the vicinity of the transition point \cite{Goldenfeld}. 
 This is related to an important property of power-laws, namely scale 
 invariance:\footnote{Power-law distributions are scale invariant in the 
 sense that the relative probability to observe an event of a given size 
 and an event ten times larger is independent of the reference scale.}
  the characteristic 
 length scale of a physical system at
its critical point is infinite, leading to self-similar, scale-free 
fluctuations. Another example is fluid turbulence, where the statistics 
of the velocity field has scale invariant properties, to a 
large extent independent of the nature of the fluid, of the power injected, etc.
\cite{Frisch}.

Power-laws are also often observed in economic and financial data
\cite{Pareto,Mandel1,Mandel2,MS,Farmer}. Compared
to
physics, however, much less effort has been devoted to 
understand these power-laws in terms of `microscopic' (i.e. agent based) models and 
to relate the
 value of the exponents to generic mechanisms. The aim of this contribution
is to give a short review of diverse power-laws observed in economy/finance,
and to discuss several simple models (most of them inspired from physics)
which naturally lead to power-laws and could serve as a starting point for 
further developments. It should be stressed that none of the models
presented here are intended to be fully realistic and complete, but are of 
pedagogical
interest: they nicely illustrate how and when power-laws can arise. This paper 
is
furthermore written in the spirit of a conference proceedings, and many rather
speculative ideas are put forward to fuel further discussions.

\section{Empirical power-laws: a short review}

\subsection{Distributional power-laws}

The oldest and most famous power-law in economy is the Pareto distribution of wealth
\cite{Pareto}.
The distribution of individual wealths $P(W)$ is often described, in its 
asymptotic tail, by a power law:
\be
P(W) \simeq \frac{W_0^\mu}{W^{1+\mu}} \qquad W \gg W_0,
\ee
where $\mu$ characterize the decay of the distribution for large $W$'s: the
smaller
the value of $\mu$, the slower the decay, and the larger the contrast between 
the
richest and the poorest. More precisely, in a Pareto population of size $N$,
 the ratio
of the largest wealth to the typical (e.g. median) wealth grows as $N^{1/\mu
}$. In the case $\mu < 1$, the average wealth diverges: this corresponds to
an economy where a finite fraction of the total wealth is in the hands of a
few individuals, even when $N \to \infty$. On the contrary, when $\mu > 1$,
the richest individual only holds a zero fraction of the total wealth
(again in the limit $N \to \infty$). Empirically, the exponent $\mu$ is in the 
range $1$ -- $2$. This Pareto tail also describes the distribution of income, 
the size of companies, of pension funds,
etc. \cite{Paretoothers,Stanley,Takayasu}.

In financial markets, the distribution of the price increments $\delta x_t=
 x(t')-x(t)$ over different time scales $t'-t$ is
important both for risk control purposes and for derivative pricing models \cite{Book}.
The availability of long series of high frequency data has motivated many 
empirical studies in the past few years.
By pooling together the statistics of a thousand U.S. stocks, it is possible to 
study quite accurately the far tail of the distribution of intra-day price 
increments $\delta x$, which can be fitted as a power law \cite{Gopi1}:
\be
P(\delta x) \simeq \frac{\delta x_0^\mu}{|\delta x|^{1+\mu}},
\ee
with $\mu$ in found to be close to $3$. Similar values have also been 
reported for Japanese stocks \cite{Gopi1}, German stocks \cite{Lux}, 
currencies \cite{Dacorogna1,Longin}, bond markets \cite{Platten}, 
and perhaps also the distribution of the (time dependent) daily volatility $\sigma$, 
defined as an average over high frequency returns \cite{Gopi2} (although other
works report a log-normal distribution \cite{Cizeau,Meyerunpub}).\footnote{Note that the tails of a log-normal distribution can be fitted
(over a restricted interval) by a power-law. Therefore it is not always easy
 to distinguish between true power-laws and effective power-laws.} This suggests that
the value of $\mu$ could be universal. Note however that the value of $\mu$ depends 
somewhat on the stock and on the period of time studied. For example, 
the value of $\mu$ for the S\&P500 during the years 1991-1995 is found to be
close to $5$. Furthermore, as the time lag used for the definition of $\delta x$
increases, the effective exponent describing the tail of the distribution increases
as the distribution becomes more and  more Gaussian like \cite{Book,Gopi1}.

\subsection{Temporal power-laws}

Actually, the price increment at time $t$ can usefully be thought of as the
product of a {\it sign part} $\epsilon_t$ and an {\it amplitude part} (or volatility) 
$\sigma_t$:
\be
\delta x_t=\epsilon_t \times \sigma_t.\label{sign}
\ee
The random variable $\epsilon_t$ has short ranged temporal correlations, 
extending over a few minutes or so on liquid markets \cite{Book,MS}. The volatility, 
on the 
contrary, has very long ranged correlations, which can be fitted as a power-law with a 
small exponent $\nu$ \cite{Memory,Ding,Dacorogna1,Cont,PCB,Cizeau}:
\be
C_1(\tau)=\langle \sigma_t \sigma_{t+\tau} \rangle - \langle \sigma_t \rangle^2 \simeq 
\frac{c_1}{\tau^\nu},\label{nudef}
\ee
with $\nu$ approximately equal to $0.1$. This behaviour is, again, seen 
on many different types of markets, and quantifies the intermittent 
activity of
these markets: volatility tends to cluster in bursts which persist 
over very
 different time scales, from minutes to months. A similar power-law behaviour 
 of the temporal correlation of the volume of transactions (number of trades) is also
observed \cite{Bonnano,Gopi2}. This is not surprising, since volatility and volume 
are strongly 
correlated. From a empirical point of view, the intermittent nature of the
activity in financial markets is similar to the energy dissipation in a
turbulent fluid \cite{Frisch,Ghasg}. 

In fact, the distribution of $\log \sigma$ is not far from being Gaussian \cite{Cizeau,
Meyerunpub}. Therefore, it 
 is natural to study the temporal correlation of $\log \sigma$ \cite{Arn,Muzy}:
\be
C_0(\tau)=\langle \log \sigma_t \log \sigma_{t+\tau} \rangle -
\langle \log \sigma_t \rangle^2.
\ee
This correlation is also found to decay very slowly with $\tau$. This decay
can be fitted by a logarithm \cite{Arn,Muzy}: $C_0(\tau)=\lambda^2 
\log(T/\tau)$, with a rather small value of $\lambda^2 \simeq 0.05$. This, together
with the assumption that $\log \sigma$ is exactly Gaussian, leads to a {\it
strict multifractal} model for the price changes \cite{Muzy}, in the sense that 
different moments of the price increments scale with different powers of time
\cite{Ghasg,Mandel3,Multif}. Within this model, one finds \cite{Muzy}:
\be
\langle |x(t+\tau) - x(t)|^q \rangle \propto \tau^{\zeta_q} \qquad \zeta_q
=\frac{q}{2}\left[1-\lambda^2 (q-2)\right],
\ee
for $\tau \ll T$ and $q\lambda^2 < 1$ (for higher values of $q$, the corresponding 
moment is divergent). It is easy to show that in this model, 
the exponent $\nu$ defined by Eq. (\ref{nudef})
is equal to $\lambda^2$. On the other hand, one can also fit 
$C_0(\tau)$ by a power-law with a small
exponent, in which case the model would only be approximately multifractal, in the
 sense that the
quantity $\langle |x(t+\tau) - x(t)|^q \rangle$ is the {\it sum} of 
different powers of $\tau$, which can also be fitted by an effective, $q$-dependent exponent 
$\zeta_q < q/2$ \cite{BPM}.

\section{Simple models of wealth distribution}

\subsection{A model with trading and speculative investment}

As a simple dynamical model of economy, one can consider a stochastic 
equation for the
wealth $W_i(t)$ of the $i^{th}$ agent at time $t$, that takes into account the
exchange of wealth between individuals through trading, and is consistent with
the basic symmetry of the problem under a change of monetary units. Since the
unit of money is arbitrary, one indeed expects that the equation governing
the evolution of wealth should be invariant when all $W_i$'s are multiplied
by
a common (arbitrary) factor. A rather general class of equation which has this 
property is the following \cite{BM,Others}:
\be
\frac{d W_i}{d t}=\eta_i(t) W_i + \sum_{j (\neq i)} J_{ij} W_j
-  \sum_{j (\neq i)} J_{ji} W_i \ , \label{fund}
\ee
where $\eta_i(t)$ is a Gaussian random variable of mean $m$ and variance 
$2\sigma^2$, which describes the 
 spontaneous growth
or decrease of wealth due to investment in stock markets, housing, etc.
The term involving the (asymmetric) matrix $J_{ij}$ describes the
amount of money that agent $j$ spends buying the production of agent $i$. 
We assume that this production is
consumable, and therefore must not be counted as part of the wealth of $i$ 
once it is
bought. The equation (\ref{fund}) is obviously invariant under the scale transformation
$W_i \to \lambda W_i$.

The simplest trading network one can think of is when all agents exchange with
all others at the same rate, i.e $J_{ij}\equiv J/N$ for all $i \neq j$. Here,
$N$ is the total number of agents, and the scaling $J/N$ is needed to make the
limit $N \to \infty$ well defined. In this case, the equation for $W_i(t)$
becomes:
\be
\frac{d W_i}{d t}=\eta_i(t) W_i + J (\overline{W} -W_i), \label{mf}
\ee
where $\overline{W}=N^{-1}\sum_i W_i$ is the average overall wealth. This is 
called, in the physics language, a `mean-field'
model since all agents feel the very same influence of their environment.
It is useful to rewrite eq. (\ref{mf}) in terms of the normalised wealths 
$w_i \equiv W_i/\overline{W}$. This leads to:
\be
\frac{d w_i}{d t}=(\eta_i(t)-m-\sigma^2) w_i + J (1 -w_i), \label{mf2}
\ee
to which one can associate the following Fokker-Planck equation for the evolution
of the density of wealth $P(w,t)$:
\be
\frac{\partial P}{\partial t}=  
\frac{\partial [J(w-1)+{\sigma^2}w]P}{\partial w}
+ {\sigma^2}
\frac{\partial}{\partial w}\left[w \frac{\partial wP}{\partial w}\right].
\label{fp}
\ee
The equilibrium, long time solution of this equation is easily shown to be:
\be
P_{eq}(w)=\frac{(\mu-1)^\mu}{\Gamma[\mu]} \frac{e^{-\frac{\mu-1}{w}}}{
w^
{1+\mu}} \qquad \mu \equiv 1 + \frac{J}{\sigma^2}.
\label{Peq}
\ee

Therefore, one finds in this model that the distribution of wealth indeed
exhibits a Pareto power-law tail for large $w$'s.  Interestingly, however, the
value of the exponent $\mu$ is {\it not universal}, and depends on the parameter
of the model ($J$ and $\sigma^2$, but not on the average growth rate $m$). 
In agreement with intuition, the exponent $\mu$  grows
(corresponding to a narrower distribution), when exchange
between agents is more active (i.e. when $J$ increases), and also when the 
success in individual investment
strategies is more narrowly distributed (i.e. when $\sigma^2$ decreases).
In this model, the exponent $\mu$ is always found to be larger than one.
This means that the wealth is not too unevenly distributed within the population.

Let us now describe more realistic trading network,
 where the number of economic neighbours to a given individual
is finite. We will assume that the matrix $J_{ij}$ is still symmetrical, and
 is either equal
 to $J$ (if $i$ and $j$ trade), or equal to $0$. A reasonable assumption is
that the graph
describing the connectivity of the population is
completely random, i.e. that two points are neighbours with probability $c/N
$ and
disconnected with probability $1-c/N$. In such a graph, 
the average number of
neighbours is equal to $c$.  We have performed some numerical simulations of
 Eq.
(\ref{fund}) for $c=4$ and have found that the wealth distribution still 
has a
power-law tail, with an exponent $\mu$ which only depends on the ratio 
$J/\sigma^2$.
However, one finds that the exponent $\mu$ can
 now be smaller than one for sufficiently small values of $J/\sigma^2$ \cite{BM}. 
 In this 
 model, one
therefore expects `wealth condensation' when the exchange rate is too small,
 in the
sense that a finite fraction of the total wealth is held by only a few 
individuals.

Although not very realistic, one could also think that the individuals
are located on the nodes of a d-dimensional
hyper-cubic  lattice, trading with their neighbours up to a finite distance.
In this case, one
knows that for $d > 2$ there exists again a phase transition between
a `social' economy where $\mu > 1$ and a rich dominated phase $\mu < 1$. On
the other hand, for $d \leq 2$,
 and for large populations, one
is always in the extreme case where $\mu \to 0$ at large times. In
the case $d=1$, i.e. operators organized along a chain-like structure (as a
 simple model of
intermediaries),
one can actually compute exactly the distribution of wealth \cite{FFH,Giardina}. 
One finds for 
example that the ratio of the maximum wealth to the typical (e.g. median)
wealth behaves as
 $\exp \sqrt{N}$, where $N$ is the size of the population,
instead of $N^{1/\mu}$ in the case of a Pareto distribution with $\mu > 0$.

The conclusion of the above
 results is that the distribution
of wealth tends to be very broadly distributed when exchanges are limited, 
either
in amplitude (i.e. $J$ too small compared to $\sigma^2$) or topologically (a
s in the above chain structure). Favoring exchanges (in particular 
with distant neighbours) seems to be an efficient way to reduce inequalities.

\subsection{Two related models}

Let us now interpret $W_i$ as the size of a company. The growth of this 
company can take place either from internal growth, depending on its 
success or failure. This leads to a term $\eta_i(t) W_i$ much as
above. Another possibility is merging with another company. If the 
merging process between two
companies is completely random and takes place at a rate $\Gamma$ per unit 
time, then the
model is exactly the same as the one considered by Derrida and Spohn \cite{DS} in the
context of `directed polymers in random media', and bears strong similarities 
with the model
discussed in the previous section. One again finds that the distribution of
$W$'s is a power-law with a non universal exponent, which depends on the values 
of $\Gamma$ and $\sigma^2$, and can be smaller than one.\footnote{In the
 language of disordered systems, this corresponds to the `glassy' phase of the 
 directed polymer, where the partition function is dominated by a few paths only.}

One can also consider a model where companies grow at a random rate $\eta$,
but may also suddenly
die at a rate $\Gamma'$ per unit time, and be replaced by a new (small) company. 
There again, one finds a stationary Pareto distribution, with a non universal 
exponent which depends continuously on $m$, $\sigma$ and $\Gamma'$.

\section{Simple models for herding and copy-cats}

We now turn to simple models for thick tails in the distribution of price 
increments in financial markets. An intuitive explanation is herding: if a 
large number of agents acting on
markets coordinate their action, the price change is likely to be huge
 due to a large unbalance between buy and sell orders \cite{Schiller}. However, this
coordination can result from two rather different mechanisms.
\begin{itemize}
\item One is the feedback of past price
changes onto themselves, which we will discuss in the following section. 
Since all agents
are influenced by the very {\it same} price changes, this can induce non trivial
collective behaviour: for example, an accidental price drop can trigger large sell 
orders, which lead to further downward moves.
\item The second is direct influence between the traders, 
through exchange of information that
leads to `clusters' of agents sharing the same decision to buy, sell, or be
inactive at any
given instant of time.
\end{itemize}

\subsection{Herding and percolation}

A simple model of how herding affects the price fluctuations was proposed in
 \cite{CB}. It assumes that the price increment $\delta x$ depends linearly on the 
 instantaneous offset
between supply and demand \cite{CB,Farmer2}. More precisely, if each operator in the market
$i$ wants to buy or sell a certain fixed quantity of the financial asset, 
one has \cite{CB}:\footnote{This can alternatively be written for the relative
price increment $\delta x/x$, which is more adapted to describe long time
scales. On short time scales, however, an additive model is preferable. See
the discussion in \cite{Book}.}
\be
\delta x=\frac{1}{\lambda} \sum_{i} \varphi_i,\label{supdem}
\ee
where $\varphi_i$ can take the values $-\, 1,\, 0$ or $+\, 1$,
depending on whether the operator $i$ is selling, inactive, or
buying, and $\lambda$ is a measure of the market depth. Note that the linearity
of this relation, even for small arguments, has been questioned by Zhang \cite{Zhang}.
Suppose now that the
 operators interact among themselves in an
heterogeneous manner: with a small probability $c/{N}$ (where
$N$ is the total number of operators on the market), two
operators $i$ and $j$ are `connected', and with probability
$1-c/{N}$, they ignore each other.  The factor $1/N$ means
that on average, the number of operator connected to any particular
one is equal to $c$ (the resulting graph is precisely the
same as the random trading graph of Section 3.1). Suppose finally that 
if two operators are
connected, they come to agree on the strategy they should follow,
i.e.\ $\varphi_i=\varphi_j$.

It is easy to understand that the population of operators clusters into groups
sharing the same opinion. These clusters are defined such that there exists
a
connection between any two operators belonging to this cluster, although the
connection can be indirect and follow a certain `path' between operators.
These clusters do not have all the same size, i.e.\ do not contain the same
number of operators. If the size of cluster ${\cal C}$ is called
$S({\cal C})$, one can write:
\be
\delta x =\frac{1}{\lambda} \sum_{\cal C} S({\cal C}) \varphi({\cal C}),
\ee
where $\varphi({\cal C})$ is the common opinion of all operators belonging 
to $\cal C$.
The statistics of the price increments $\delta x$ therefore reduces to 
the statistics
of the size of clusters, a classical problem in percolation theory 
\cite{Stauffer}. One finds that as long as $c < 1$ (less than one `neighbour' 
on 
average with whom one can exchange
information), then all $S({\cal C})$'s are small compared to the total 
number of traders
$N$. More precisely, the distribution of cluster sizes takes the following
form in the limit where  $1-c=\epsilon \ll 1$:
\be
P(S) \propto_{S \gg 1} \frac{1}{S^{5/2}} \exp -\epsilon^2 S \qquad S \ll {N}
.\label{PN}
\ee
When $c=1$ (percolation threshold), the distribution becomes a pure
power-law with an exponent $\mu=3/2$, and the Central Limit Theorem tells us
that in this case, the distribution of the price increments $\delta x$
is precisely a pure symmetrical L\'evy distribution of index $\mu=3/2$ \cite{Book}
(assuming that $\varphi=\pm \, 1$ play identical roles, that is if
there is no global bias pushing the price up or down). If $c < 1$, on
the other hand, one finds that the L\'evy distribution is truncated
exponentially, leading to a larger effective tail exponent $\mu$ \cite{CB}. If $c > 1$,
a finite fraction of the ${N}$ traders have the same opinion: this leads to
a crash.

\subsection{Avalanches of opinion changes}

The above model is interesting but has one major drawback: one has to
assume that the parameter $c$ is smaller than one, but relatively
close to one such that Eq.\ (\ref{PN}) is valid, and non trivial
statistics follows. One should thus explain why the value of $c$
spontaneously stabilises in the neighbourhood of the critical value
$c=1$. Furthermore, this model is purely static, and does not specify
how the above clusters evolve with time. As such, it cannot address the
problem of volatility clustering. Several extensions of this simple model
have been proposed \cite{Stauffer2,Iori}, in particular to increase the value
of $\mu$ from $\mu=3/2$ to $\mu \sim 3$ and to account for volatility clustering. 

One particularly interesting model
is inspired by the
recent work of Dahmen and Sethna \cite{Sethna}, that describes the behaviour
 of random
magnets in a time dependent magnetic field. Transposed to the present problem 
(as first suggested in \cite{Book})
, this model describes the collective behaviour of a set of traders 
exchanging information, but
having all different {\it a priori} opinions. One trader can however
change his mind and take the opinion of his neighbours if the coupling
is strong, or if the strength of his {\it a priori} opinion is
weak. More precisely, the opinion $\varphi_i(t)$ of agent $i$ at time $t$ is
 determined as:
\be
\varphi_i(t)=\mbox{sign}\left(h_i(t)+\sum_{j=1}^{N} J_{ij}\varphi_j(t)\right
),
\ee
where $J_{ij}$ is a connectivity matrix describing how strongly agent $j$
affects agent $i$, and $h_i(t)$ describes the {\it a priori} opinion of agent 
$i$:
$h_i >0$ means a propensity to buy, $h_i <0$ a propensity to sell.
We assume that $h_i$ is a random variable with a time dependent mean 
$\overline{h}(t)$
and root mean square $\Delta$. The quantity $\overline{h}(t)$ represents for
example confidence in long term economy growth ($\overline{h}(t)>0$), 
or fear of recession ($\overline{h}(t)<0$, leading to
a non zero average pessimism or optimism.  Suppose that one starts at $t
=0$ from a
`euphoric' state, where $\overline{h} \gg \Delta,J$, such that $\varphi_i=
1$
for all $i$'s.\footnote{Here $J$ denotes the order of magnitude of $\sum_{j}
 J_{ij}$} Now, confidence is decreased progressively. How will sell orders appear ?

Interestingly, one finds that for small enough influence 
(or strong heterogeneities
of agents' anticipations), i.e. for $J \ll \Delta$, the average opinion $O(t)=\sum_i 
\varphi_i(t)/N$
evolves continuously from $O(t=0))=+1$ to $O(t \to \infty)=-1$. The situation 
changes
when imitation is stronger since a discontinuity then appears in $O(t)$ around 
a `crash' time
$t_c$, when a {\it finite} fraction of the population simultaneously change opinion. 
The gap $O(t_c^-)-O(t_c^+)$ opens continuously as $J$ crosses a critical
value $J_c(\Delta)$ \cite{Sethna}.
For $J$ close to $J_c$, one finds that the sell
orders again organise as avalanches of various sizes, distributed as a
power-law with an exponential cut-off. In the `mean-field' case where $J_{ij
}\equiv J/N$
for all ${ij}$, one finds $\mu=5/4$. Note that in this case, the value of the 
exponent $\mu$ {\it is universal}, and does not depend, for example, on the shape of
 the distribution of the $h_i$'s, but only on some global properties of the
connectivity matrix $J_{ij}$. 

A slowly oscillating $\overline{h}(t)$ can therefore lead
to a succession of bull and bear markets, with a strongly non Gaussian, 
intermittent behaviour, since most of the activity is concentrated 
around the crash times $t_c$. Some modifications of this model are however
needed to account for the empirical value $\mu \sim 3$ observed on the distribution
of price increments (see the discussion in \cite{Stauffer2}). 

Note that the same model can be applied 
to other interesting situations, for example to describe how applause end in a
concert hall \cite{Rava} (here, $\varphi=\pm 1$ describes, respectively, a clapping and 
a quiet person, and $O(t)$ is the total clapping activity).
Clapping can end abruptly (as often observed, at least by the present author) 
when imitation is strong, or smoothly when many fans are 
present in the audience. A static version of the same model has been proposed 
to describe rational
group decision making \cite{Galam}.

\section{Models of feedback effects on price fluctuations}

\subsection{Risk-aversion induced crashes}

The above average `stimulus' $\overline{h}(t)$ may also strongly depend on the 
past behaviour of the price itself. For example, past positive trends are,
 for many investors, incentives to buy, and vice-versa. Actually, for a 
 given trend amplitude,
 price drops tend to feedback more strongly on investors' behaviour than price 
 rises. Risk-aversion creates an asymmetry between positive and negative
 price changes \cite{Langevin}. This is reflected by option markets, and 
 pushes the price of out-of-the-money puts up. 
 
 Similarly, past periods of high volatility increases the risk of 
 investing in stocks, and usual portfolio theories
then suggest that sell orders should follow. A simple mathematical transcription of 
these effects is to write Eq. (\ref{supdem}) in a linearized, continuous 
time form:\footnote{In the following, the herding effects described
by $J_{ij}$ are neglected, or more precisely, only their {\it average effect}
encoded by $\overline h$ is taken into account.}
\be
\frac{dx}{dt} \equiv u=\frac{1}{\lambda} \overline{h}(t),
\ee
and write a dynamical equation for $\overline{h}(t)$ which encodes the above
 feedback
effects \cite{Langevin,Farmer2}:
\be
\frac{d\overline{h}}{dt} = a u - bu^2 - c\overline{h} + \eta(t),
\ee
where $a$ describes trends following effects, $b$ risk aversion effects (breaking the
$u \to -u$ symmetry), $c$
 is a mean
reverting term which arises from market clearing mechanisms, and $\eta$ is a
 noise
term representing random external news \cite{Langevin}. Eliminating 
$\overline{h}$ from the above equations leads to:
\be
\frac{du}{dt}= - \gamma u - \beta u^2 +
\frac{1}{\lambda} \eta(t) \equiv -\frac{\partial V}{\partial u}
+  \frac{1}{\lambda} \eta(t) \label{Langevin1}
\ee
where $\gamma=c-a/\lambda$ and $\beta=b/\lambda$. Equation (\ref{Langevin1}) 
represents the evolution of the position $u$ of a viscous fictitious particle
in a `potential' $V(u)=\gamma u^2/{2} + {\beta} u^3/{3}$. If trend following 
effects are not too strong, $\gamma$ is positive and $V(u)$ has a local 
minimum for $u=0$, and a local maximum for $u^*=- \gamma/\beta$, beyond
 which the potential plummets to $-\infty$.\footnote{If $\gamma$ is negative, 
 the minimum appears for a positive value of the return $u^*$. This
 corresponds to a speculative bubble. See \cite{Langevin}.} A `potential barrier' 
 $V^*={\gamma u^{*2}}/{6}$ separates the (meta-)stable region around $u=0$ 
 from the unstable region. The nature of the motion of $u$ in such a 
 potential is the following: starting
at $u=0$, the particle has a random harmonic-like motion in the vicinity of 
$u=0$ until an `activated' event (i.e. driven by the noise term) brings 
the particle near
$u^*$. Once this barrier is crossed, the fictitious particle reaches 
$-\infty$ in finite time. In financial terms,
the
regime where $u$ oscillates around $u=0$ and where $\beta$ can be neglected, 
is the `normal' fluctuation regime. This normal regime
can however be interrupted by `crashes', where the time derivative of the
price
becomes very large and negative, due to the risk aversion term $b$ which
destabilizes the price by amplifying the sell orders. The interesting point is that 
these two regimes can be
clearly separated since the average time $t^*$ needed for such crashes
to occur is exponentially large in $V^*$ \cite{Chandrasekar}, and can thus appear
only very rarely. A very long time scale is therefore naturally generated in this model.

Note that in this line of thought, a crash
occurs because of an improbable succession
of unfavorable events, and not due to a single large event in particular.
Furthermore, there are
no `precursors' in this model:
 before $u$ has reached $u^*$, it is impossible to decide whether it
will do so or whether it will quietly come back in the `normal' region $u
\simeq 0$. Solving the Fokker-Planck equation associated to the Langevin 
equation
(\ref{Langevin1}) leads to a stationary state with a power law tail for the
distribution of $u$ (i.e. of the instantaneous price increment) decaying as
$u^{-2}$
for $u \to -\infty$.
More generally, if the risk aversion term took the form $-b u^{1+\mu}$, the
 negative
tail would decay as $u^{-1-\mu}$.

\subsection{Dynamical volatility models}

The simplest model that describes volatility feedback effects is to write
an {\sc arch} like equation \cite{arch}, which relates today's activity to a measure
of yesterday's activity, for example:
\be
\sigma_{k}=\sigma_{k-1} + K (\sigma_0-\sigma_{k-1}) + g |\delta x_{k-1}|,
\ee
where $\sigma_0$ is an average volatility level, $K$ a mean-reverting term,
and $g$ describes how much yesterday's observed price change affects the behaviour
of traders today. Since $|\delta x_{k-1}|$ is a noisy version of $\sigma_{k-1}$,
the above equation is, in the continuous time limit, a Langevin equation with 
{\it multiplicative} noise:
\be
\frac{d\sigma}{d t}= K(\sigma_0-\sigma) + g\sigma\eta(t) \label{mf3},
\ee
which is, up to notation changes, exactly the same equation as (\ref{mf2}) above. 
Therefore, the stationary distribution of the volatility in this model is again
given by Eq. (\ref{Peq}), with the tail exponent now given by $\mu-1 \propto K/g^2$:
over-reactions to past informations (i.e. large values of $g$) decreases the tail
exponent $\mu$. Therefore, one again finds a non universal exponent in this model, 
which is bequeathed to the distribution of price increments if one assumes that the
`sign' contribution to $\delta x_k$ (see Eq. (\ref{sign})) has thin tails.

Note that the temporal correlations of the volatility $\sigma$ can be exactly 
calculated within this model \cite{Graham}, and is found to be exponentially decaying, at variance
with the slow power-law (or logarithmic) decay observed empirically. Furthermore,
the distribution (\ref{Peq}) does not concur with the nearly log-normal distribution
of the volatility that seems to hold empirically \cite{Cizeau,Meyerunpub}. 

At this point, the slow decay of the volatility can be ascribed to two rather
different mechanisms. One is the existence of traders with many different 
time horizons, as suggested in \cite{harch,PCB}. If traders are affected not only
by yesterday's price change amplitude $|\delta x_{k-1}|$, but also by price
changes on coarser time scales $|x_{k}-x_{k-p}|$, then the correlation function
is expected to be a sum of exponentials with decay rates given by $p^{-1}$. 
Interestingly, if the different $p$'s are uniformly distributed on a log scale,
the resulting sum of exponentials is to a good approximation decaying as
a logarithm. More precisely:\footnote{This mechanism is well known in the physics of
slow, glassy systems, where the relaxation times $p$ are the exponential of some local 
activation energy $E$. At low temperatures, any small dispersion of $E$ will generate a $1/p$ 
distribution for $p$ over a wide time interval, and eventually to a logarithmic relaxation.} 
\be
C(\tau)=\frac{1}{\log(p_{\max}/p_{\min})} 
\int_{p_{\min}}^{p_{\max}} d(\log p) \exp(-\tau/p)
\simeq  \frac{\log(p_{\max}/\tau)}{\log(p_{\max}/p_{\min})},
\ee
whenever ${p_{\min}}\ll \tau \ll {p_{\max}}$.
Now, the human time scales are indeed in a natural pseudo-geometric progression: hour, 
day, week, month, trimester, year \cite{PCB}. 

Yet, some recent numerical simulations of traders allowed to switch between different 
strategies (active/inactive, or chartist/fundamentalist) suggest
strongly intermittent behaviour \cite{Lux2,Iori,GBM}, and a slow decay of the volatility
correlation function without the explicit existence of logarithmically distributed
time scales. An intuitive, semi-analytical explanation of this numerical finding is however still
lacking. Note that from a purely phenomenological point of view, one can define 
a model that
assumes that the volatility today is the retarded result of past influences:
\be
\sigma_k = \sum_{p=0}^\infty M(p) \eta_{k-p},\label{retarded}
\ee
where the $\eta$'s are uncorrelated shocks and $M(p)$ a memory kernel describing 
how much the past is remembered. If one chooses a power-law decay for $M(p)$ with
an exponent $\alpha$, then the decay of the correlation function of $\sigma$ is also
a power-law with an exponent $\nu$ (defined in Eq. (\ref{nudef})) given by 
$\nu=2 \alpha -1$. The value $\nu=0$ corresponds to a memory kernel decaying as
$1/\sqrt{p}$, i.e., as the probability of not returning to the origin for a
random walk of length $p$. Whether this is a mere coincidence is left for future
investigations. Note however that a decomposition such as (\ref{retarded}) naturally
leads to a normal distribution for $\sigma$, very different from the empirical 
log-normal behaviour of the volatility. A consistent market model leading simultaneously to
a nearly log-normal distribution and a nearly logarithmic (in time) decay of the 
volatility correlation remains to be built. In this respect, the cascade construction of
Mandelbrot et al. \cite{Mandel3} does indeed have these two properties exactly, but is 
non-causal (the volatility today depends on future events) and lacks an intuitive 
interpretation.

\section{Concluding remarks}

Many ideas have been presented in this rather hairy paper. Most of them will perhaps 
turn out to be wrong, but will hopefully motivate some further work to understand
the origin of power-law {\it distributions} and power-law {\it correlations} 
in financial time series. From an empirical point of view, the exponents describing
the tails of the price increments distribution and the decay of the volatility 
correlations are rather robust and suggest some kind of universality, probably related
to the fact that all speculative markets obey common rules where simple 
human psychology (greed and fear) coupled to basic mechanisms of price formation ultimately 
lead to the emergence of scaling and power-laws. Still, many points remain obscure:
we have seen above that models that appear naturally in the context of economy and finance
contain multiplicative noise, which is a simple mechanism for power-law distributions (as
emphasized in, e.g. \cite{Others,Stauffer2}). However, these models 
generically lead to {\it
non universal exponents} (as discussed above in the context of the Pareto tails) that depends continuously on the value of the parameters.  
Recent 
progress in the empirical study of the volatility (using, e.g., wavelets \cite{Arn,Muzy})
suggests that the volatility results from some sort of multiplicative cascade, as postulated
in \cite{Mandel3}. A convincing `microscopic' (i.e. trader based) model that explains 
this observation would at this stage be very valuable, and would shed light on the 
possible relevance of the pseudo geometric human time scales on the decay of the
volatility correlations.

\section*{Acknowledgments} I want to warmly thank all my collaborators for sharing with me
their knowledge: P. Cizeau, R. Cont, I. Giardina, L. Laloux, A. Matacz, M. M\'ezard, 
M. Meyer and 
M. Potters. Several discussions with E. Bacry, R. da Silvera, M. Dacorogna, D. Farmer, 
T. Lux, R. Mantegna, J.F. Muzy, G. Stanley, 
D. Stauffer and S. Solomon have been very useful.

\end{document}